\documentstyle[aps,graphicx]{revtex}
\newcommand{\inieq}{\begin{eqnarray}}            
\newcommand{\fineq}{\end{eqnarray}}            
\newcommand{\diff}{{\rm\,d}}                    

\def\p{\mbox{\boldmath $p$}}

\def\q{\mbox{\boldmath $q$}}

\def\k{\mbox{\boldmath $k$}}

\begin{document}
\title{Neutrino-Nucleus Quasi-Elastic Scattering in a Relativistic Model}
\author{C.~Giusti, A.~Meucci, and F.~D.~Pacati}
\address{Dipartimento di Fisica Nucleare e Teorica, Universit\`{a} degli 
Studi di Pavia,\\ 
Istituto Nazionale di Fisica Nucleare, Sezione di Pavia, I-27100 
Pavia, Italy}
\maketitle

\begin{abstract}
A relativistic distorted-wave impulse-approximation model is applied to 
neutral-current and charged-current quasi-elastic neutrino-nucleus scattering. 
The effects of final state interactions are investigated and the sensitivity 
of the results to the strange nucleon form factors is discussed in view of 
their possible experimental determination 
\end{abstract}

\section{Introduction}
Neutrino-nucleus scattering has gained in recent years a wide interest that 
goes beyond the study of the intrinsic properties of neutrinos and extends to
different fields, such as astrophysics, cosmology, particle and nuclear physics.
The observation of neutrino oscillations and the proposal and realization of
new experiments, aimed at determining neutrino properties with high accuracy,
renewed interest in neutrino scattering on complex nuclei. In fact, neutrino
detectors contain nuclei and a detailed knowledge of the $\nu$-nucleus
interaction is necessary for a proper interpretation of the experimental data. 
Neutrino-nucleus scattering, however, is not only an useful tool to detect
neutrinos, but plays an important role also in understanding various
astrophysical processes. The influence of neutrinos extends to cosmological
questions. Moreover, neutrinos provide a suitable tool to test the limits of the
standard model, the properties of the weak interaction and for investigating 
nuclear structure. In hadronic and nuclear physics neutrinos can give 
information on the structure of the hadronic weak current and on the strange 
quark contribution to the spin structure of the nucleon. 

Thus, neutrino physics is of great interest and involves many different
phenomena. The problem is that neutrinos are elusive particles. They are
chargeless, almost massless, and only weakly interacting. Their presence can
only be inferred detecting the particles they create when colliding or 
interacting with matter. Nuclei are often used as neutrino detectors providing
relatively large cross sections. Therefore, the interpretation of data requires 
reliable calculations where nuclear effects are properly taken into account. 

General review papers about neutrino-nucleus interactions can be found in
\cite{Walecka,peccei,alb,kolbe03}.
Both weak neutral-current (NC) and charged-current (CC) scattering have 
stimulated detailed analyses in the intermediate-energy region 
\cite{min,engel,kim,singh,hayes,volpe,alberico,kolbe,jacho,kim96,umino,co1} 
using a variety of methods, including Fermi Gas (FG), Random-Phase-Approximation 
(RPA) and Shell-Model (SM) calculations. The effects of Final State Interactions 
(FSI) were investigated within the Relativistic FG (RFG) model \cite{bleve}, the 
RPA \cite{gar93} and in the continuum RPA (CRPA) theory \cite{Botrugno}. Nuclear 
structure effects on the determination of the strange quark contribution in NC 
scattering were studied in \cite{alberico,barbaro}, and in \cite{vdv} in the 
framework of a Relativistic Plane Wave Impulse Approximation (RPWIA). The effects 
of FSI on the ratio of proton-to-neutron cross sections in NC scattering were 
discussed in \cite{alberico,hor,gar92,Martinez}.

We study CC and NC $\nu$- and $\bar\nu$-nucleus scattering in the QE 
region. In this region the dominant contribution is given by one-nucleon 
knockout processes, where the interaction occurs on a nucleon, that is bound in 
the nucleus but is assumed to be a quasi-free nucleon in the process, this 
nucleon is emitted and the remaining nucleons are spectators. In the QE region 
we have applied the same Relativistic Distorted Wave Impulse Approximation 
(RDWIA) model that was successfully tested in comparison with data for the 
exclusive (e,e$'$p) knockout reaction \cite{book,meucci1}. The analysis of NC 
$\nu$-nucleus reactions, however, introduces additional complications, as the 
final neutrino cannot be measured in practice and a final hadron has to be 
detected: the corresponding cross sections are therefore semi-inclusive in the 
hadronic sector and inclusive in the leptonic one. The same approach is here 
applied to the CC scattering where only the outgoing nucleon is detected. The 
case of the inclusive CC scattering 
where only the outgoing charged lepton is detected was studied in \cite{cc} 
through a relativistic Green's function approach, that was firstly applied to 
the inclusive QE electron scattering \cite{ee} and where 
FSI are accounted for by means of a complex optical potential but without loss 
of flux. 

The formalism is outlined in Sec. 2. Nuclear effects, in particular the effects
of FSI, are discussed in Sec. 3. The effects of the strange nucleon form factors 
and their possible determination are investigated Sec. 4. Some conclusions are 
drawn in Sec. 5.

\section{Formalism of Quasi-Elastic Neutrino-Nucleus Scattering}
\label{sec.for}

The $\nu$($\bar\nu$)-nucleus cross section for the process where one nucleon is 
emitted is given, in the one-boson exchange approximation, by the contraction 
between the lepton and the hadron tensor, i.e.,
\begin{eqnarray}
\diff \sigma = \frac {G_{\mathrm{F}}^2 } {2} \ 2\pi \
 L^{\mu\nu}\ W_{\mu\nu}\ \frac {\diff^3k} {(2\pi)^3} \ 
 \frac {\diff^3p_{\mathrm N}} {(2\pi)^3} \ ,
\label{eq.cs1}
\end{eqnarray}
where $G_{\mathrm{F}} \simeq 1.16639 \times 10^{-11}$ MeV$^{-2}$ is the Fermi 
constant,  $k^\mu_i = (\varepsilon_i,\k_i)$, $k^\mu = (\varepsilon,\k)$ are the 
four-momentum of the incident and final leptons, respectively, and 
$\p_{\mathrm N}$ is the momentum of the emitted nucleon. For CC processes 
$G_{\mathrm{F}}^2$ has to be multiplied 
by $\cos ^2\vartheta_{\mathrm C} \simeq 0.9749$, where 
$\vartheta_{\mathrm C}$ is the Cabibbo angle.

The lepton tensor $L^{\mu\nu}$ has a similar structure as in electron scattering
and separates into a symmetrical and an antisymmetrical part \cite{book,cc,nc}. 
The components of the lepton tensor are kinematical factors which depend only 
on the lepton kinematics. 
The components of the hadron tensor are given by bilinear products of the 
transition matrix elements of the nuclear weak-current operator $J^{\mu}$ between
the initial state $|\Psi_0\rangle$ of the nucleus, of energy $E_0$, and the 
final states, of energy $E_{\textrm {f}}$, that are given by the product of a 
discrete (or continuum) state $|n\rangle$ of the residual nucleus and a 
scattering state $\chi^{(-)}_{\p_{\mathrm N}}$ of the emitted nucleon, with 
momentum $\p_{\mathrm N}$  and energy $E_{\mathrm N}$: 
\begin{eqnarray}
W^{\mu\nu}(\omega,q) & = & 
 \sum_{n}  \langle n;\chi^{(-)}_{\p_{\mathrm N}} 
\mid J^{\mu}(\q) \mid \Psi_0\rangle 
\langle 
\Psi_0\mid J^{\nu\dagger}(\q) \mid n;\chi^{(-)}_{\p_{\mathrm N}}\rangle 
\nonumber \\
& \times & \delta (E_0 +\omega - E_{\textrm {f}}) \ ,
\label{eq.ha1}
\end{eqnarray}
where the sum runs over all the states of the residual nucleus.  
In the first order perturbation theory and using the impulse approximation, 
the transition amplitude is assumed to be adequately described as the sum of 
terms similar to those appearing in the exclusive (e,e$'$p) knockout reaction 
\cite{book,meucci1}
\begin{equation}
\langle n;\chi^{(-)}_{\p_{\mathrm N}}\mid J^{\mu}(\q) \mid \Psi_0\rangle = 
\langle\chi^{(-)}_{\p_{\mathrm N}}\mid   j^{\mu}
(\q)\mid \varphi_n \rangle  \ . \label{eq.amp}
\end{equation} 
The transition amplitudes are thus obtained in a one-body representation and
contain three ingredients: the one-body nuclear weak current $j^{\mu}$,
the one-nucleon overlap function $\varphi_n = \langle n | \Psi_0\rangle$, 
that is a single-particle (s.p.) bound state wave function, and the s.p.
scattering wave function $\chi^{(-)}$ for the outgoing nucleon.

Bound and scattering states are consistently derived in the model as 
eigenfunctions of an optical potential. In practice, calculations are performed 
with the same phenomenological ingredients already used in the RDWIA 
calculations for the (e,e$'$p) reaction. 
The s.p overlap functions $\varphi_n$ are Dirac-Hartree solutions of a 
relativistic Lagrangian, containing scalar and vector potentials. They are 
obtained in the framework of the relativistic mean field theory and reproduce
the s.p. properties of several nuclei\cite{adfx,lala}. 
The relativistic scattering wave function is written in terms of its upper
component, following the direct Pauli reduction scheme and solving 
a Schr\"odinger-like equation containing equivalent central and 
spin-orbit potentials, written in terms of the relativistic scalar and 
vector potentials \cite{clark,HPa}. Calculations have been performed with the 
energy-dependent and A-dependent EDAD1 optical potential of \cite{chc}. 

The s.p. operator related to the weak current is
\begin{eqnarray}
  j^{\mu} &=&  F_1^{\textrm V}(Q^2) \gamma ^{\mu} + 
             i\frac {\kappa}{2M} F_2^{\textrm V}(Q^2)\sigma^{\mu\nu}q_{\nu}	  
	     -G_{\textrm A}(Q^2)\gamma ^{\mu}\gamma ^{5}     
 \ \ ({\mathrm {NC}}) 
\ , \nonumber 
	     \\  	     
  j^{\mu} &=&  \Big[F_1^{\textrm V}(Q^2) \gamma ^{\mu} + 
             i\frac {\kappa}{2M} F_2^{\textrm
	     V}(Q^2)\sigma^{\mu\nu}q_{\nu}	  
\nonumber \\  &-&G_{\textrm A}(Q^2)\gamma ^{\mu}\gamma ^{5} + 
 F_{\textrm P}(Q^2)q^{\mu}\gamma ^{5}\Big]\tau^{\pm} \  \ ({\mathrm {CC}}) \ ,  
	     \label{eq.nc}
\end{eqnarray}
where $\tau^{\pm}$ are the isospin operators, $\kappa$ is the anomalous part of 
the magnetic moment, $q^{\mu} = (\omega , \q)$, with $Q^2 = |\q|^2 - \omega^2$, 
is the four-momentum transfer, and
$\sigma^{\mu\nu}=\left(i/2\right)\left[\gamma^{\mu},\gamma^{\nu}\right]$.
$G_{\textrm A}$ is the axial form factor and $F_{\textrm P}$ is the induced 
pseudoscalar form factor. 
The weak isovector Dirac and Pauli form factors, $F_1^{\textrm V}$ and
$F_2^{\textrm V}$, are related to the corresponding electromagnetic form
factors by the conservation of the vector current (CVC) hypothesis 
\cite{Walecka} plus, for NC reactions, a possible isoscalar strange-quark 
contribution $F_i^{\mathrm s}$, i.e., 
\begin{eqnarray}
F_i^{\mathrm {V,p(n)}} &=& \left(\frac{1}{2} - 
2\sin^2{\theta_{\mathrm W}}\right)
 F_i^{\mathrm {p(n)}} -\frac{1}{2} F_i^{\mathrm {n(p)}} - 
 \frac{1}{2} F_i^{\mathrm s} \ 
 \ \ \ \ ({\mathrm {NC}}) \ , \nonumber \\
 F_i^{\mathrm V} &=& 
 F_i^{\mathrm p} - F_i^{\mathrm n} 
 \ \ \ \ ({\mathrm {CC}}) \ , \label{eq.nc1}
\end{eqnarray}
where $\theta_{\mathrm W}$ is the Weinberg angle 
$(\sin^2{\theta_{\mathrm W}} \simeq 0.23143)$. The electromagnetic form factors 
are taken from \cite{bba} and the strange form factors as \cite{alb}
\begin{eqnarray}
F_1^{\mathrm s}(Q^2) =  \frac {(\rho^{\mathrm s} + 
\mu^{\mathrm s}) \tau}{(1+\tau) (1+Q^2/M_{\mathrm V}^2)^2}\ , \ 
F_2^{\mathrm s}(Q^2) =  \frac {\left(\mu^{\mathrm s}-\tau \rho^{\mathrm s}  
\right)}{(1+\tau) (1+Q^2/M_{\mathrm V}^2)^2}\ ,
\label{eq.sform}
\end{eqnarray}
where $\tau = Q^2/(4M^2)$ and $M_{\mathrm V}$ = 0.843 GeV. The constants  
$\rho^{\mathrm s}$ and $\mu^{\mathrm s}$ describe the strange quark
contribution to the electric and magnetic form factors, respectively.
The axial form factor is expressed as \cite{mmd}
\begin{eqnarray}
G_{\mathrm A} &=& \frac{1}{2} \left( \tau_3 g_{\mathrm A} - 
g^{\mathrm s}_{\mathrm A}\right) G\ \ \ \ ({\mathrm {NC}}) \ , \nonumber \\
G_{\mathrm A} &=&   g_{\mathrm A} G\ \ \ \ ({\mathrm {CC}}) \ , \label{eq.ga}
\end{eqnarray}
where $g_{\mathrm A} \simeq 1.26$, $g^{\mathrm s}_{\mathrm A}$ describes 
possible strange-quark contributions,  $G = (1+Q^2/M_{\mathrm A}^2)^{-2}$, and 
$\tau_3 = +1 (-1)$ for proton (neutron) knockout.
The axial mass has been taken as $M_{\mathrm A}$ = (1.026$\pm$0.021) GeV 
\cite{bernard}.

The single differential cross section with respect to the outgoing  nucleon 
kinetic energy $T_{\mathrm N}$ is obtained after integrating over the energy and
angle of the final lepton and over the solid angle of the final nucleon.  

In the calculations a pure SM description is assumed for nuclear structure. The
state $n$ is assumed to be a one-hole state in the SM and  $\varphi_n$ are s.p. 
SM states with a unitary spectral strength. The sum over in 
Eq. (\ref{eq.ha1}) runs over all the occupied states in the SM. In this 
way we include the contributions of all the nucleons in the nucleus but 
neglect the effects of correlations that, anyhow, are expected to be small 
in the situations considered in the present investigation.

The cross section for the $\nu$($\bar\nu$)-nucleus scattering where only
one-nucleon is detected is obtained from the sum of all the integrated 
exclusive one-nucleon knockout channels. FSI are described by means of a 
complex optical potential whose imaginary part gives an absorption that reduces 
the calculated cross section. It accounts for the flux lost in a particular 
channel and that goes towards other channels. This approach is conceptually 
correct for an exclusive reaction, where only one channel contributes, but it 
would be conceptually wrong for an inclusive reaction, where all the channels 
contribute and the total flux must be conserved. In fact, for the inclusive 
electron scattering \cite{ee} and for the CC scattering where only the 
outgoing lepton is detected \cite{cc} we adopt a different treatment of FSI, 
which makes use of a complex optical potential and where the total flux is 
conserved.  Here, we consider semi-inclusive situations where an emitted 
nucleon is always detected and some of the reaction channels which are 
responsible for the imaginary part of the optical potential, like fragmentation 
of the nucleus, re-absorption, etc., are not included in the experimental 
cross section. From this point of view, it is correct to include the absorptive 
imaginary part of the optical potential. There are, however, contributions that 
are not included in our model and that can be included in the experimental cross 
section, for instance, contributions due to multi-step processes, where the 
outgoing nucleon is re-emitted after re-scattering in a detected channel 
simulating the kinematics of a QE reaction. The relevance of these 
contributions depends on kinematics and should not be too large in the 
situations considered in this paper. Anyhow, even if the use of an optical 
potential with an absorptive imaginary part can introduce some uncertainties 
in the comparison with data, we deem it a more correct and clearer way to 
evaluate the effects of FSI. 

\section{Nuclear Effects and Final State Interactions}
Calculations have been performed for NC and CC  $\nu_\mu$ ($\bar \nu_{\mu}$)
scattering from $^{12}$C in an energy range between 500 and  1000 MeV, where 
one-nucleon knockout is expected to be the most important contribution. In this 
Section nuclear effects are investigated in calculations where the strange form 
factors are neglected. The effects of the strange nucleon form factors are 
discussed in the next Section. 

\begin{figure}[ht]
\begin{center}
\includegraphics[height=90mm,width=90mm]{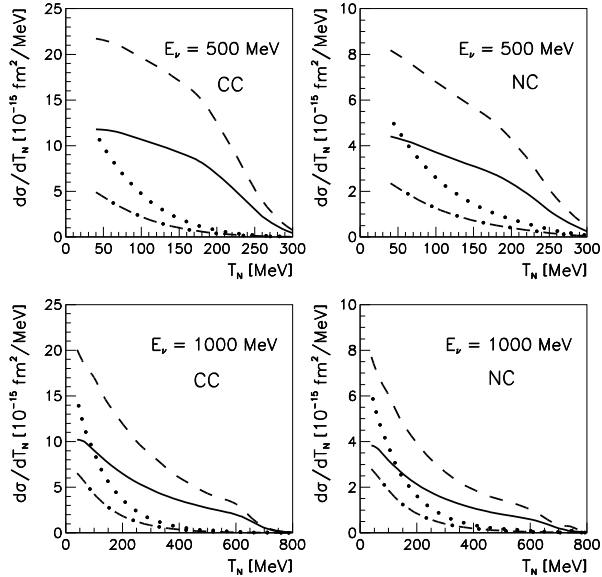}
\vspace{2mm}\caption{Differential cross sections of the CC and NC $\nu_\mu$ 
($\bar \nu_{\mu}$) QE scattering on $^{12}$C as a function of 
T$_{\mathrm N}$. Solid and dashed lines are the results in RDWIA and RPWIA, 
respectively, for an incident neutrino. Dot-dashed and dotted lines are the 
results in RDWIA and RPWIA, respectively, for an incident antineutrino. The 
strangeness contribution in the NC scattering is neglected \label{fig1}}
\end{center}
\end{figure}

Nuclear effects are included in the phenomenological ingredients for the bound 
and scattering states. Calculations performed with different bound state wave 
functions and with different optical potentials are not very sensitive to the 
choice and to the details of the phenomenological ingredients. Large effects 
are, however, produced by FSI. An example is shown in Figure~\ref{fig1}, where 
the cross sections of the $^{12}$C$\left(\nu_{\mu},\mu^-p\right)$ and 
$^{12}$C$\left(\bar\nu_{\mu},\mu^+n\right)$ CC reactions and of the  
$^{12}$C$\left(\nu_{\mu},\nu_{\mu} p\right)$ and 
$^{12}$C$\left(\bar\nu_{\mu},\bar\nu_{\mu} p\right)$ NC reactions are compared
in RPWIA and RDWIA at $E_{\nu (\bar \nu)}= 500$ and $1000$ MeV. FSI reduce the 
cross sections of  $\simeq 50$\%. This reduction is due to the imaginary 
part of the optical potential and is in agreement with the reduction found in 
the (e,e$'$p) calculations. We note that the cross sections for 
an incident neutrino are larger than for an incident antinuetrino.

\section{Strange Nucleon Form Factors}
It is well known that the net strangeness of the 
nucleon is zero. It is also known, however, that according to the quantum field 
theory in the cloud of a physical nucleon there must be pairs of strange 
particles. From the viewpoint of QCD the nucleon consists of $u$ and $d$ quarks 
and of a sea of $q{\bar q}$ pairs produced by virtual gluons. Then, the
question is: how do the sea quarks, in particular strange quarks, contribute 
to the observed properties of the nucleon? The first evidence that the constant 
$g^{\mathrm s}_{\mathrm A}=G^{\mathrm s}_{\mathrm A}(Q^2=0)$, that
characterizes the matrix element of the axial strange current, is different from
zero and large was found  by the EMC experiment at CERN \cite{EMC}, in 
a measurement of deep inelastic scattering of polarized muons on polarized
protons. This result triggered new experiments and a lot of theoretical work.
It is very important that different and alternative methods are used to
determine the matrix elements of the strange current. NC $\nu$ scattering is one
of these methods and a suitable tool to investigate $g^{\mathrm s}_{\mathrm A}$.

Different nucleon form factors contribute to the s.p. weak current operator of 
the NC scattering of Eq. (\ref{eq.nc}). A combination of different 
measurements is required for a complete information. The electromagnetic form 
factors, $F_1$ and $F_2$ in Eq. (\ref{eq.nc1}), can be investigated in electron 
scattering. The value of the Weinberg angle $\theta_{\mathrm W}$ can be 
obtained from measurements of NC processes. Quasi-elastic CC
scattering can give information on the axial form factor $G_{\mathrm A}$, whose
determination is very important in general and in particular if we want to
determine $g^{\mathrm s}_{\mathrm A}$, that is highly correlated to 
$G_{\mathrm A}$ and thus to the axial mass $M_{\mathrm A}$. The
strange form factors, $F_1^{\mathrm s}$, $F_1^{\mathrm s}$, and 
$G^{\mathrm s}_{\mathrm A}$, can be investigated in NC $\nu$ scattering and in 
Parity-Violating Electron Scattering (PVES). PVES is essentially sensitive to 
$F_1^{\mathrm s}$ and $F_1^{\mathrm s}$ or, equivalently, to the strange
electric and magnetic from factors $G^{\mathrm s}_{\mathrm E}$ and 
$G^{\mathrm s}_{\mathrm M}$.  A determination of $G^{\mathrm s}_{\mathrm A}$
in PVES is hindered by radiative corrections. In contrast, NC $\nu$ scattering 
is primarily sensitive to $G^{\mathrm s}_{\mathrm A}$. The interference with 
the strange vector form factors can be resolved by complementary experiments of 
PVES.  

\begin{figure}[ht]
\begin{center}
\includegraphics[height=90mm,width=90mm]{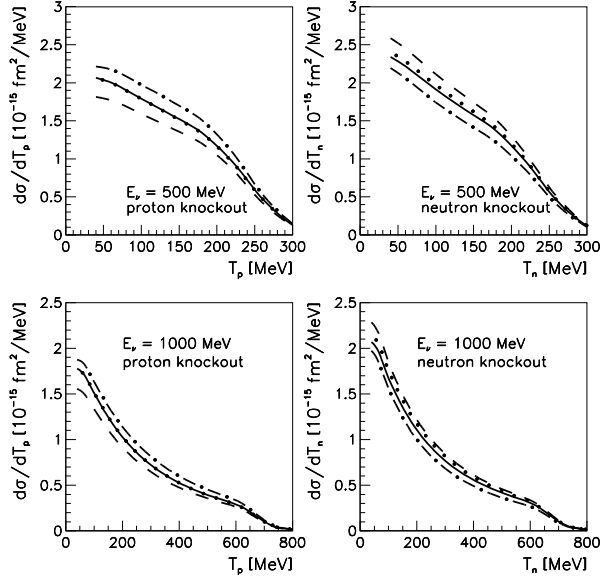}
\vspace{2mm}\caption{Differential cross sections of the NC $\nu_{\mu}$ QE
scattering on $^{12}$C as a function of $T_{\mathrm N}$. Dashed lines are the 
results with no strangeness contribution, 
solid lines with $g^{\mathrm s}_{\mathrm A} = -0.10$, dot-dashed lines
with $g^{\mathrm s}_{\mathrm A} = -0.10$ and $\mu^{\mathrm s} = -0.50$,
dotted lines with $g^{\mathrm s}_{\mathrm A} = -0.10$ and 
$\rho^{\mathrm s} = +2$. \label{fig2}}
\end{center}
\end{figure}

A determination of the form factors is beyond the scope of the present 
investigation. Our main aim here is to study the sensitivity of NC 
$\nu$-nucleus scattering to the strange quark contribution. In Figure~\ref{fig2} 
the cross sections calculated, both for proton and neutron emission, with a 
particular choice for the values of the parameters, 
$g^{\mathrm s}_{\mathrm A}=-0.10$, $\mu^{\mathrm s}=-0.50$, and 
$\rho^{\mathrm s}=+2$,  are compared with the results obtained without 
strange form factors. The cross sections with 
$g^{\mathrm s}_{\mathrm A} = -0.10$ are enhanced in the case of proton knockout 
and reduced in the case of neutron knockout by $\simeq 10$\% 
with respect to those with $g^{\mathrm s}_{\mathrm A} = 0$ . 
The effect of $\mu^{\mathrm s}$ is comparable to that of 
$g^{\mathrm s}_{\mathrm A}$, whereas the contribution of $\rho^{\mathrm s}$ is 
very small for neutron knockout and practically negligible for proton knockout.  
\begin{figure}[ht]
\begin{center}
\includegraphics[height=90mm,width=90mm]{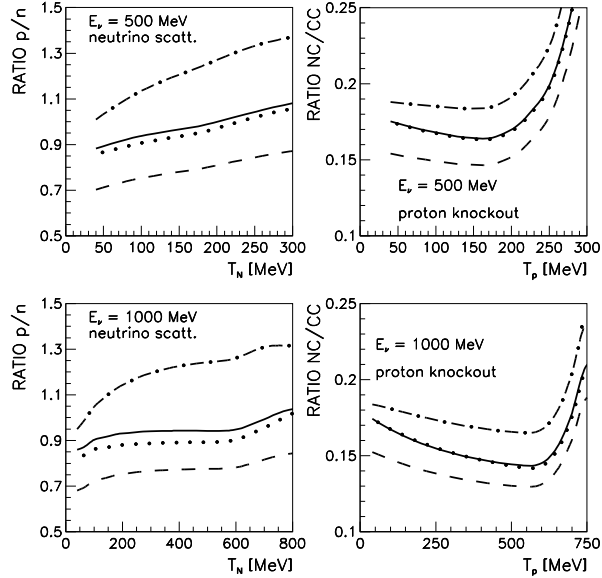}
\vspace{2mm}\caption{Ratio of proton-to-neutron NC cross sections (left panel) and of
NC-to-CC cross sections (right panel) of the $\nu$ QE scattering
on $^{12}$C.  Line convention as in Figure~\ref{fig2}. \label{fig3}}
\end{center}
\end{figure}

An absolute cross section measurement is a very hard experimental task due to 
difficulties in the determination of the neutrino flux. Thus, ratios of cross 
sections were proposed as an alternative way to extract 
$g^{\mathrm s}_{\mathrm A}$. Difficulties due to the determination of the 
absolute neutrino flux are reduced in the ratios. Moreover, also nuclear 
effects can be strongly reduced in the ratios. The effects of FSI are large on 
the cross sections and almost negligible in the ratios, where they give a 
similar contribution to the numerator and to the denominator \cite{strange}. 
In contrast, strangeness effects can be emphasized in the ratios, where form 
factors may contribute in a different way, for instance with a different sign, 
in the numerator and in the denominator. 

Two different ratios are presented in Figure~\ref{fig3}.  The ratio of 
proton-to-neutron (p/n) NC cross sections is sensitive to the 
strange-quark contribution as the interference between 
$g^{\mathrm s}_{\mathrm A}$ and $g_{\mathrm A}$ contributes with an opposite 
sign in the numerator and in the denominator [see Eq. (\ref{eq.ga})]. A precise 
measurement of this ratio appears, however, problematic due to the difficulties 
associated with neutron detection. A measurement of the ratio of the NC-to-CC 
(NC/CC) cross sections appears more feasible and will be measured at 
FINeSSE \cite{fin}. Although sensitive to strangeness only in the numerator, 
the NC/CC ratio is simply related to the number of events with an outgoing 
proton and a missing mass with respect to the events with an outgoing proton 
in coincidence with a muon.  The ratios in Figure~\ref{fig3} are sensitive to 
$g^{\mathrm s}_{\mathrm A}$  and $\mu^{\mathrm s}$, while the effects  of 
$\rho^{\mathrm s}$ are very small. The results show similar features at 
different energies of the incident neutrino.

\section{Conclusions}

We have presented RDWIA calculations for CC and NC $\nu$($\bar\nu$)-nucleus QE 
scattering. The effects of FSI are large on the cross section and almost 
negligible in the (p/n) and (NC/CC) ratios. The results obtained with the 
strange form factors are sensitive to $g^{\mathrm s}_{\mathrm A}$  and 
$\mu^{\mathrm s}$ and practically insensitive to $\rho^{\mathrm s}$. 
Measurements of the (p/n) and (NC/CC) ratios would be interesting to 
determine the constant $g^{\mathrm s}_{\mathrm A}$ that characterizes the 
matrix element of the axial strange current. The interference with the strange 
vector form factors can be resolved with complementary experiments of PVES.

\end{document}